\documentclass[aps,pre,groupedaddress,showpacs,preprint]{revtex4}
\usepackage{graphicx}
\usepackage[dvips]{epsfig}
\usepackage{dcolumn}
\usepackage{subfigure}%
\usepackage{bm}
\usepackage{amssymb}
\usepackage{amsmath}

\begin{document}
\title{Geometrical approach to tumor growth}

\author{Carlos Escudero}

\affiliation{Mathematical Institute, University of Oxford, 24-29 St Giles', Oxford OX1 3LB, United Kingdom}

\begin{abstract}
Tumor growth has a number of features in common with a physical process known as molecular beam epitaxy. Both growth
processes are characterized by the constraint of growth development to the body border, and surface diffusion of cells/particles at the growing edge. However, tumor growth implies an approximate spherical symmetry that makes necessary a
geometrical treatment of the growth equations. The basic model was introduced in a former article
[C. Escudero, Phys. Rev. E {\bf 73}, 020902(R) (2006)], and in the present work we extend our analysis and try to shed light on the possible geometrical principles
that drive tumor growth. We present two-dimensional models that reproduce the experimental observations, and analyse the unexplored three-dimensional case, for which new conclusions on tumor growth are derived.
\end{abstract}

\pacs{87.10.+e, 68.35.Fx, 87.19.-j}
\maketitle

\section{Introduction}

One of the highest mortality rates in developed countries is due to cancer. For this reason, it is one of the most studied diseases, and however, it is still far from being well understood. Aside from the intensive medical and biological research, an increasing number of theoretical models is being introduced in order to describe some of the fundamental properties of tumors. One of the most common mathematical approaches is the use of partial differential equations, which has lead to some interesting results in the field~\cite{castro}.

A different methodology was used by Br\'{u} {\it et al.}~\cite{bru1,bru2}, that employed some tools of fractal geometry, as scaling analysis, to characterize the rough interface of growing solid tumors. They found strong empirical evidence that a broad class of tumors belong to the same universality class: the molecular beam epitaxy (MBE) universality class. MBE is a well known process in physics, in which a crystal surface grows due to the external input of atoms coming from a beam directed to the growing surface~\cite{barabasi}. This finding allowed them to postulate that tumor dynamics has some features in common with MBE, say: (a) a linear growth rate, (b) the constraint of growth activity to the outer border of the tumor/crystal, and (c) surface diffusion at the growing edge. All of these features were again tested against experiment with a positive result.

These experimental observations lead to a new and very interesting picture of tumor growth. The usual assumed exponential growth is replaced by a linear growth, and the dynamics is constrained to the peripheral region, because it is assumed that what happens in the core of the tumor has little effect on growth. Surface diffusion has been identified as a mechanism for favoring tumor growth. The host tissue exerts pressure on solid tumors which opposes their growth, but surface diffusion drives the tumor cells to the concavities of the interface, keeping the number of neighboring cells that belong to the host tissue to a minimum. Since these cells are responsible of the pressure exerted on the tumor, surface diffusion minimizes the pressure on the interface, and favors the propagation of the tumor~\cite{bru1,bru2}.

This new theoretical description of a solid tumor was used to develop a strategy to stop the growth~\cite{bru3}. Suggested by the above findings, Br\'{u} {\it et al.} proposed that an enhancement of the pressure on the tumor surface would be able to decelerate, and eventually to stop tumor growth. They performed a new experiment in which they observed the response of the tumor to an enhancement of the inmune response. An increase of the number of neutrophils shifted the dynamics of the interface from the MBE universality class to the much slower quenched Edwards-Wilkinson (QEW) universality class, and to the pinning of the tumor interface~\cite{bru3}. This technique was later applied to a patient with a terminal cancer, who subsequentially improve his state till finally achieve a good health, possibly due to the applied treatment~\cite{bru4}.

This success has a fundamental importance for the development of efficient therapies, and therefore it would be highly desirable to achieve a good theoretical understanding of the models used. MBE dynamics is described by the Mullins-Herring equation~\cite{mullins,herring}
\begin{equation}
\label{mh}
\partial_t h=-K\nabla^4 h + F + \eta({\bf x},t),
\end{equation}
where $h$ is the interface height, $K$ is the surface diffusion coefficient, and $\eta({\bf x},t)$ is a Gaussian noise
with zero mean and correlations given by
\begin{equation}
<\eta({\bf x},t)\eta({\bf x'},t')>=D\delta({\bf x}-{\bf x'})\delta(t-t').
\end{equation}
This equation was developed to describe crystal growth, so it assumes that the substrate is planar and does not change its size in time. This no longer applies to tumors, since they are approximately spherical and grow linearly in time; both features are possibly the main discrepancies between standard MBE dynamics and the dynamics of tumors~\cite{bru2}. The same criticism applies to the equation describing QEW dynamics~\cite{barabasi}
\begin{equation}
\label{qew}
\partial_t h=K\nabla^2 h + F + \eta({\bf x},h),
\end{equation}
where $\eta({\bf x},h)$ is a quenched disorder with zero mean and correlations given by
\begin{equation}
\label{qew2}
<\eta({\bf x},h)\eta({\bf x'},h')>=D\delta({\bf x}-{\bf x'})\Delta(h-h'),
\end{equation}
and the function $\Delta$ characterizes the nature of the quenched disorder.

In a former article, we developed a stochastic partial differential equation describing the same dynamics as Eq.(\ref{mh}) but with the correct geometrical symmetries~\cite{escudero}. The analysis of this equation revealed that it was able to reproduce some of the fundamental mechanisms of tumor growth found in experiments~\cite{bru1,bru2}. In the present work we extend the geometrical approach to tumor growth by developing and analyzing spherically symmetric equations describing both MBE and QEW dynamics. These equations model the behaviour found experimentally in the $(1+1)-$dimensional
case~\cite{bru1,bru2,bru3}, and allow us to predict what would happen in the more realistic and unexplored case of
$(2+1)-$dimensional geometry. We also derive the equations using a more systematic technique, that allows us to conjecture what are the geometrical principles that drive tumor growth.

\section{Expansion from a potential}
\label{potential}

In general, a planar stochastic growth equation may be written in the form
\begin{equation}
\frac{\partial h(x,t)}{\partial t}=G[h(x,t)]+\eta(x,t),
\end{equation}
where $G$ is the deterministic growth term and $\eta$ is the noise.
If we want the mean field equation to describe a conservative dynamics (as for instance a diffusion), then
the deterministic part must have the form of a continuity equation
\begin{equation}
\frac{\partial h(x,t)}{\partial t}=-\nabla \cdot j(x,t),
\end{equation}
where the macroscopic current $j(x,t)$ describes the flux of cells on the surface. The current $j(x,t)$ arises in general from differences in the local pressure $p(x,t)$ as argued before, following the law
\begin{equation}
j(x,t)=\nabla \Pi(x,t),
\end{equation}
where $\Pi$ is defined as a pressure potential. We can perform the expansion of $\Pi$ in terms of the pressure
\begin{equation}
\Pi(x,t)=-A_1 p(x,t)+A_2 \nabla^2 p(x,t)+\cdots.
\end{equation}
Since the difference in pressure comes mainly from the differences in height of the different parts of the interface, we may assume that
$p \propto h$. Finally, we can write the deterministic part of the evolution equation as
\begin{equation}
\frac{\partial h(x,t)}{\partial t}= A_1 \nabla^2 h - A_2 \nabla^4 h + \dots,
\end{equation}
where we can identify the terms present in the drifts of both Eq.(\ref{mh}) and Eq.(\ref{qew}).

The equation of growth of a general Riemannian surface reads
\begin{equation}
\partial_t \vec{r}({\bf s},t)=\hat{n}({\bf s},t)\Gamma[\vec{r}({\bf s},t)]+\vec{\Phi}({\bf s},t),
\end{equation}
where the $d+1$ dimensional surface vector $\vec{r}({\bf s},t)=\{r_\alpha({\bf s},t)\}_{\alpha=1}^{d+1}$ runs over the surface as ${\bf s}=\{s^i\}_{i=1}^d$ varies in a parameter space (in the following, latin indices vary from 1 to $d$ and greek indices from 1 to $d+1$). In this equation $\hat{n}$ stands for the unitary vector normal at the surface at $\vec{r}$, $\Gamma$ contains a deterministic growth mechanism that causes growth along the normal $\hat{n}$ to the surface, and $\vec{\Phi}$ is a random force acting on the surface. In our case the deterministic part should include a term modelling cell diffusion in the tumor border. When surface diffusion occurs to minimize the surface area the corresponding term in the equation is~\cite{marsili}:
\begin{equation}
\label{surdiff}
\Gamma_s=-K\Delta_{BL}H,
\end{equation}
where $\Delta_{BL}$ is the Beltrami-Laplace operator
\begin{equation}
\Delta_{BL}=\frac{1}{\sqrt{g}}\partial_i(\sqrt{g}g^{ij}\partial_j),
\end{equation}
$g_{ij}$ is the metric tensor and g is its determinant, $\partial_i=\partial/\partial s^i$ is a covariant derivative, and $H=\hat{n} \cdot \Delta_{BL} \vec{r}$ is the mean curvature.
Summation over repeated indices is always assumed along this work.
Finally, the unitary normal vector is given by
$\hat{n}=g^{-1/2}\partial_1\vec{r}\times\cdots\times\partial_d\vec{r}$. The equation of growth can be derived straightforwardly from here~\cite{escudero}, but since our aim is to understand the geometrical principles underlying tumor growth, we write the more general growth equation for those
cases in which the drift can be derived from a potential:
\begin{equation}
\partial_t \vec{r}(s,t)=-\frac{1}{\sqrt{g(s)}}\frac{\delta \mathcal{V}[\vec{r}(s,t)]}{\delta \vec{r}(s,t)}.
\end{equation}
In our case the potential $\mathcal{V}$ depends on the mean curvature $H$ of the interface. This dependence can be expressed
in a power series expansion:
\begin{equation}
\label{expansion}
\mathcal{V}=\int d^ds \sqrt{g}\sum_{i=0}^N K_i H^i=\sum_{i=0}^N \mathcal{V}_i.
\end{equation}
From here we can derive straightforwardly the stationary probability distribution functional $P[r(s,t)]$, which yields the probability of the interface configuration $r(s,t)$ in the limit $t \to \infty$
\begin{equation}
P[r(s,t)]=\mathcal{N}\exp \left[ -\frac{\mathcal{V}[r(s,t)]}{D/2} \right],
\end{equation}
where $D$ is the noise strength and $\mathcal{N}$ is the normalization constant. In this work, we are more interested in the dynamical rather than in the stationary properties of the model, so we will focus on the stochastic partial differential equation which describes tumor growth.
In the general equation, the contribution to the drift reads
\begin{equation}
\Gamma_i=-\frac{1}{\sqrt{g}}\hat{n} \cdot \frac{\delta \mathcal{V}_i}{\delta \vec{r}}=
K_i \left(H^{i+1}-i\Delta_{BL}H^{i-1}-iH^{i-1}\sum_{j=1}^d \lambda_j^2 \right),
\end{equation}
where $\lambda_j$ are the eigenvalues of the matrix of the coefficients of the second fundamental form and express the principal curvatures of the surface~\cite{marsili}.

\section{Stochastic equations for tumor growth}

In this section we will build the models for describing the growth of a non-treated tumor. First of all, we will derive the
zeroth, first, and second order terms in the expansion Eq.(\ref{expansion}) in one and two dimensions. The corresponding contributions to the drift in polar coordinates read~\cite{polar}
\begin{subequations}
\begin{eqnarray}
\Gamma_0(d=1) &=& \frac{1}{r^2}\frac{\partial^2 r}{\partial \theta^2}, \\
\Gamma_1(d=1) &=& 0, \\
\Gamma_2(d=1) &=& -\frac{1}{r^4} \frac{\partial^4 r}{\partial \theta^4}, \\
\Gamma_0(d=2) &=& \frac{1}{r^2}\left(\frac{\partial^2 r}{\partial \theta^2}
+\frac{1}{\sin^2(\theta)}\frac{\partial^2 r}{\partial \phi^2}\right), \\
\Gamma_1(d=2) &=& \frac{2}{r^3}\left(\frac{\partial^2 r}{\partial \theta^2}+
\frac{1}{\sin^2(\theta)}\frac{\partial^2 r}{\partial \phi^2}\right), \\
\Gamma_2(d=2) &=& -\frac{1}{r^4}\left(\frac{\partial^4 r}{\partial \theta^4}+
\frac{2}{\sin^2(\theta)}\frac{\partial^4 r}{\partial \theta^2 \partial \phi^2}+
\frac{1}{\sin^4(\theta)}\frac{\partial^4 r}{\partial \phi^4}\right),
\end{eqnarray}
\end{subequations}
where we have linearized the different derivatives of $r$ about zero and chosen only the most relevant terms in the renormalization group sense. Retaining only the linear terms is a valid approximation whenever sharp changes in the
tumor interface are absent~\cite{escudero}. It is very important to realize that $\Gamma_1(d=1)$ vanish identically,
since $H=\lambda_1$ in $d=1$. This is a consequence of the Gauss-Bonnet theorem, which states that the integral of the Gaussian curvature $K$ on a closed surface is a constant. Since $H=K$ in $d=1$, the variation of $\mathcal{V}_1$ is zero.
Another important fact is that $\Gamma_1(d > 1)$ is strictly nonlinear in the derivatives of $r$ in the case of a planar geometry, so it will never survive after a linearization~\cite{marsili}. We will show what are the consequences of this in the next sections.

In the case of non-treated tumors, only the $\Gamma_2$ term seems to appear in the dynamics. In this case, we can derive the following equations describing the tumor interface dynamics in $(1+1)-$dimensions
\begin{equation}
\label{tumor2d}
\frac{\partial r}{\partial t}= -\frac{K}{r^4} \frac{\partial^4 r}{\partial \theta^4} + F + \frac{1}{\sqrt{r}}\eta(\theta,t),
\end{equation}
where the noise $\eta(\theta,t)$ is Gaussian, with zero mean, and correlation given by
\begin{equation}
<\eta(\theta,t)\eta(\theta',t')>=D\delta(\theta-\theta')\delta(t-t'),
\end{equation}
and $(2+1)-$dimensions
\begin{equation}
\label{tumor3d}
\frac{\partial r}{\partial t}= -\frac{K}{r^4}\left(\frac{\partial^4 r}{\partial \theta^4}+
\frac{2}{\sin^2(\theta)}\frac{\partial^4 r}{\partial \theta^2 \partial \phi^2}+
\frac{1}{\sin^4(\theta)}\frac{\partial^4 r}{\partial \phi^4}\right)+ F +
\frac{1}{r\sqrt{|\sin(\theta)|}}\eta(\theta,\phi,t),
\end{equation}
where the noise $\eta(\theta,\phi,t)$ is Gaussian, with zero mean, and correlation given by
\begin{equation}
<\eta(\theta,\phi,t)\eta(\theta',\phi',t')>=D\delta(\theta-\theta')\delta(\phi-\phi')\delta(t-t'),
\end{equation}
and the noise \emph{must} be interpreted in the It\^{o} sense. A more detailed derivation of these equations can be found in Ref. \cite{escudero}.

To analyze these equations we will perform a small noise expansion~\cite{sancho}, where the solution is decomposed as follows:
\begin{equation}
r(\theta,t)=R(t)+\sqrt{D}\rho(\theta,t),
\end{equation}
in the $(1+1)-$dimensional case, where $R$ is the deterministic solution, given by $R(t)=Ft+R_0$. The stochastic perturbation obeys the equation
\begin{equation}
\frac{\partial \rho}{\partial t}=-\frac{K}{(R_0+Ft)^4}\frac{\partial^4 \rho}{\partial \theta^4}+\frac{1}{\sqrt{R_0+Ft}}\eta(\theta,t),
\end{equation}
where the noise $\eta(\theta,t)$ is Gaussian, with zero mean, and correlation given by
\begin{equation}
<\eta(\theta,t)\eta(\theta',t')>=\delta(\theta-\theta')\delta(t-t').
\end{equation}
The discrete Fourier transformed version of this equation reads
\begin{equation}
\frac{d\rho_n}{dt}=-\frac{Kn^4}{(R_0+Ft)^4}\rho_n+\frac{1}{\sqrt{R_0+Ft}}\eta_n(t),
\end{equation}
where the noise $\eta_n(t)$ is Gaussian, with zero mean, and correlation given by
\begin{equation}
<\eta_n(t)\eta_m(t')>=(2\pi)^{-1}\delta_{n,-m}\delta(t-t'),
\end{equation}
where $\delta_{n,-m}$ denotes the Kronecker symbol. The mean value of the stochastic process obeys the equation
\begin{equation}
\frac{d<\rho_n>}{dt}=-\frac{Kn^4}{(R_0+Ft)^4}<\rho_n>,
\end{equation}
that can be solved to yield
\begin{equation}
<\rho_n(t)>=\exp\left(\frac{Kn^4}{3F}\left[\frac{1}{(R_0+Ft)^3}-\frac{1}{R_0^3}\right]\right)<\rho_n(0)>.
\end{equation}
We can also calculate the two-point correlation function using standard techniques~\cite{doering}
\begin{eqnarray}
\nonumber
d<\rho_n(t)\rho_m(t)>=<\rho_n(t+dt)\rho_m(t+dt)>-<\rho_n(t)\rho_m(t)>= \\
\nonumber
<d\rho_n(t)\rho_m(t)>+<\rho_n(t)d\rho_m(t)>+ <d\rho_n(t)d\rho_m(t)>= \\
-\frac{K}{(R_0+Ft)^4}(n^4+m^4)<\rho_n(t)\rho_m(t)>dt+\frac{1}{R_0+Ft}\delta_{n,-m}(2\pi)^{-1}dt,
\end{eqnarray}
where we have used the facts that $\eta_n(t)=(2\pi)^{-1/2}dW_n(t)/dt$ and $dW_n(t)^2=dt$, and where $dW_n(t)$ denotes the increment of a Wiener process. And thus, the two-point correlation function $C_{n,m}(t)=<\rho_n(t)\rho_m(t)>$ obeys the equation
\begin{equation}
\frac{dC_{n,m}(t)}{dt}=-\frac{K}{(R_0+Ft)^4}(n^4+m^4)C_{n,m}(t)+\frac{(2\pi)^{-1}}{R_0+Ft}\delta_{n,-m}.
\end{equation}
The exact solution of this equation is
\begin{eqnarray}
\nonumber
C_{n,m}(t)=\exp \left[-\frac{K(n^4+m^4)}{3F} \left(\frac{1}{R_0^3}-\frac{1}{(R_0+Ft)^3}\right) \right]\times \\
\nonumber
\left[C_{n,m}(0)+\frac{\delta_{n,-m}}{6\pi F}\exp \left[\frac{K(n^4+m^4)}{3FR_0^3}\right]\times \right. \\
\left.
\left(\mathrm{Ei}\left[-\frac{K(n^4+m^4)}{3FR_0^3}\right]-\mathrm{Ei}\left[-\frac{K(n^4+m^4)}{3F(R_0+Ft)^3}\right]\right)
\right],
\end{eqnarray}
where $\mathrm{Ei}$ denotes the exponential integral.

In the $(2+1)-$dimensional case we can perform the expansion
\begin{equation}
r(\theta,\phi,t)=R(t)+\sqrt{D}\rho(\theta,\phi,t),
\end{equation}
where $R(t)=R_0+Ft$. The stochastic perturbation obeys the equation
\begin{equation}
\frac{\partial \rho}{\partial t}= -\frac{K}{(R_0+Ft)^4}\left(\frac{\partial^4 \rho}{\partial \theta^4}+
\frac{2}{\sin^2(\theta)}\frac{\partial^4 \rho}{\partial \theta^2 \partial \phi^2}+
\frac{1}{\sin^4(\theta)}\frac{\partial^4 \rho}{\partial \phi^4}\right) +
\frac{1}{(R_0+Ft)\sqrt{|\sin(\theta)|}}\eta(\theta,\phi,t),
\end{equation}
which implies
\begin{equation}
\frac{d\rho_{n,m}}{dt}=-\frac{K}{(R_0+Ft)^4}\left( n^4+\frac{8}{3}n^2m^2+\frac{8}{3}m^4 \right)\rho_{n,m}+\frac{A}{R_0+Ft}\eta_{n,m}(t),
\end{equation}
where $A=160F(\pi/4|2)/(63\pi)$, and $F$ denotes an incomplete elliptic integral of the first kind~\cite{elliptic}. The noise is again Gaussian, with zero mean, and correlation given by
\begin{equation}
<\eta_{n,m}(t)\eta_{p,q}(t')>=(2\pi)^{-2}\delta_{n,-p}\delta_{m,-q}\delta(t-t').
\end{equation}
We can derive the equation for the
first moment
\begin{equation}
\frac{d<\rho_{n,m}>}{dt}=-\frac{K}{(R_0+Ft)^4}\left( n^4+\frac{8}{3}n^2m^2+\frac{8}{3}m^4 \right)<\rho_{n,m}>,
\end{equation}
and solve it to obtain
\begin{equation}
<\rho_{n,m}(t)>=\exp \left(-\frac{K(8m^4+8m^2n^2+3n^4)}{9F}\left[\frac{1}{R_0^3}-\frac{1}{(R_0+Ft)^3}\right] \right)<\rho_{n,m}(0)>.
\end{equation}
The equation for the two-point correlation function can be straightforwardly derived and is
\begin{eqnarray}
\nonumber
\frac{dC_{n,m,p,q}(t)}{dt}=-\frac{K}{(R_0+Ft)^4}\left(n^4+\frac{8}{3}n^2m^2+\frac{8}{3}m^4+p^4+\frac{8}{3}p^2q^2+\frac{8}{3}q^4\right)C_{n,m,p,q}(t) \\
+\frac{B}{(R_0+Ft)^2}\delta_{n,-p}\delta_{m,-q},
\end{eqnarray}
where $B=6400F(\pi/4|2)^2/(3969\pi^4) (\approx 0.03)$, and $C_{n,m,p,q}(t)=<\rho_{n,m}(t)\rho_{p,q}(t)>$. The solution to this equation is
\begin{eqnarray}
\nonumber
C_{n,m,p,q}(t)=\exp \left[-\frac{K}{9F}(8m^4+8m^2n^2+3n^4+3p^4+8p^2q^2+8q^4)\left(\frac{1}{R_0^3}-\frac{1}{(R_0+Ft)^3} \right)\right] \\
\nonumber
\times
\left[ C_{n,m,p,q}(0)+B\delta_{n,-p}
\delta_{m,-q}\exp\left(\frac{K}{9FR_0^3}(8m^4+8m^2n^2+3n^4+3p^4+8p^2q^2+8q^4)\right)\right.\times \\
\left.
\int_0^t \exp\left[-\frac{K(8m^4+8m^2n^2+3n^4+3p^4+8p^2q^2+8q^4)}{9F(F\tau+R_0)^3}\right](F\tau+R_0)^{-2}d\tau
\right].
\end{eqnarray}

One can see that our exact solutions reveal some interesting characteristics of the growth. The mean value of the small perturbation decreases in time proving the stability of the mean field radially symmetric solution, both in one and two dimensions. The correlation functions give a much more interesting information. The correlations generated by the noise are much smaller in the two-dimensional case: not only the numerical prefactor is smaller, but in the one-dimensional case these correlations decrease in time as an exponential integral, while in the two-dimensional case the decay is much faster. This means that the effect of the noise is much stronger in one dimension, while the deterministic dynamics are more robust in two dimensions, something that might have serious consequences in tumor therapy, as we will show below.

\section{Growth in a disordered medium}

In the models commented in the introduction, it seems that the disorder is induced by the enlarged amount of neutrophils introduced to stop tumor growth. But actually, the medium in which tumors grow is highly disordered.Tumors are extensively infiltrated by immune cells which may constitute as much as one third of its volume. Both the tumor phenotype and the tumor environment are very heterogeneous. The former is the result of accumulating random mutations, variable enviromental selection forces and perhaps restriction of proliferate capacity in non-stem cell components of the tumor. In addition, the tumor environment is extremely heterogeneous primarily due to disordered angiogenesis and blood flow. These facts suggest that one possible explanation for the noise term appearing in the equations of the last section comes from the underlying disorder. In fact, if we consider Eq.(\ref{qew}) and the corresponding correlations of the quenched disorder Eq.(\ref{qew2}), we see that for large $F$ the function $h$ will behave as $h \sim Ft$, and assuming that the function $\Delta$ models short-range correlations (as found experimentally~\cite{bru3}), implies $\Delta(h-h') \propto \delta(t-t')$. This means that, far from the pinning threshold, the role of the disorder is to induce thermal fluctuations in the dynamics as those found in Eq.(\ref{mh}). Physically, this means that a rapidly moving interface samples so many values of the disorder in a small time interval, that the overall effect is that of a time dependent noise. This implies in turn that the intensity of the noise is proportional to the disorder, and thus an enhancement in the inmune response corresponds to a stronger noise in Eqs.(\ref{tumor2d}, \ref{tumor3d}), while the tumor is still in a state far enough from the pinning threshold.

Not only the noise, but also the diffusion terms vary from Eq.(\ref{mh}) to Eq.(\ref{qew}). We can understand this effect if we suppose the terms in the expansion Eq.(\ref{expansion}) dependent on the disorder. Assuming that $K_0 \sim D$ and that $K_2$ is independent of $D$, where $D$ is the intensity of the disorder, then we find that the $K_0$ term is the most relevant in high disorder, while $K_2$ would be the most important term in other case. We can use these facts to build a more general model of tumor growth in the $(1+1)-$dimensional setting
\begin{equation}
\label{gtumor2d}
\frac{\partial r}{\partial t}= \frac{K_0}{r^2}\frac{\partial^2 r}{\partial \theta^2}-\frac{K_2}{r^4} \frac{\partial^4 r}{\partial \theta^4} + F + \frac{1}{\sqrt{r}}\eta(\theta,r),
\end{equation}
where the correlations of the quenched disorder are given by
\begin{equation}
<\eta(\theta,r)\eta(\theta',r')>=\delta(\theta-\theta')\Delta(r-r').
\end{equation}
In weak disorder, this equation reduces to Eq.(\ref{tumor2d}), because the disorder behaves as a thermal noise and the term proportional to $K_0$ loses its importance, as explain above.

In the $(2+1)-$dimensional case the situation is different. In opposition to the former case, now the term proportional to $K_1$ does not vanish identically as shown in section~\ref{potential}. Taking into account this fact, we can again build the more general equation for tumor growth, that reads
\begin{eqnarray}
\nonumber
\frac{\partial r}{\partial t}=\frac{K_0}{r^2}\left(\frac{\partial^2 r}{\partial \theta^2}
+\frac{1}{\sin^2(\theta)}\frac{\partial^2 r}{\partial \phi^2}\right)+\frac{2K_1}{r^3}\left(\frac{\partial^2 r}{\partial \theta^2}+\frac{1}{\sin^2(\theta)}\frac{\partial^2 r}{\partial \phi^2}\right) \\
-\frac{K_2}{r^4}\left(\frac{\partial^4 r}{\partial \theta^4}+\frac{2}{\sin^2(\theta)}\frac{\partial^4 r}{\partial \theta^2 \partial \phi^2}+ \frac{1}{\sin^4(\theta)}\frac{\partial^4 r}{\partial \phi^4}\right) + F +
\frac{1}{r\sqrt{|\sin(\theta)|}}\eta(\theta,\phi,r),
\label{gtumor3d}
\end{eqnarray}
where the correlations of the quenched disorder are given by
\begin{equation}
<\eta(\theta,\phi,r)\eta(\theta',\phi',r')>=\delta(\theta-\theta')\delta(\phi-\phi')\Delta(r-r').
\end{equation}

Now, it is our goal to understand what differences appear in the evolution of the equations for tumor growth in different dimensions, mainly due to the presence of the $K_1$ term in the $(2+1)-$dimensional model.
For this, we will study the deterministic counterparts of Eqs.(\ref{gtumor2d}, \ref{gtumor3d}). In the case of $(1+1)-$dimensions we have
\begin{equation}
\frac{\partial r}{\partial t}= \frac{K_0}{r^2}\frac{\partial^2 r}{\partial \theta^2}-\frac{K_2}{r^4} \frac{\partial^4 r}{\partial \theta^4} + F,
\end{equation}
which admits the radially symmetric solution $r(t)=R_0+Ft$. If we perform the linear stability analysis of this solution by adding a small perturbation $\rho$, we obtain the equation
\begin{equation}
\frac{d\rho_n}{dt}=-\left[\frac{K_0n^2}{(R_0+Ft)^2}+\frac{K_2n^4}{(R_0+Ft)^4}\right]\rho_n,
\end{equation}
that can be solved to yield
\begin{equation}
\label{sol2d}
\rho_n(t)=\exp\left(\frac{K_2n^4}{3F}\left[\frac{1}{(R_0+Ft)^3}-\frac{1}{R_0^3}\right]+\frac{K_0n^2}{F}\left[\frac{1}{R_0+Ft}-\frac{1}{R_0}\right]\right)\rho_n(0).
\end{equation}
This reveals that all the Fourier modes $n \neq 0$ of the solution are linearly stable for $t>0$. The $n=0$ mode is marginal, but this is unimportant because it implies perturbations homogeneous in $\theta$. On the other hand, the equation corresponding to the $(2+1)-$dimensional case reads
\begin{eqnarray}
\nonumber
\frac{\partial r}{\partial t}=\frac{K_0}{r^2}\left(\frac{\partial^2 r}{\partial \theta^2}
+\frac{1}{\sin^2(\theta)}\frac{\partial^2 r}{\partial \phi^2}\right)+\frac{2K_1}{r^3}\left(\frac{\partial^2 r}{\partial \theta^2}+\frac{1}{\sin^2(\theta)}\frac{\partial^2 r}{\partial \phi^2}\right) \\
-\frac{K_2}{r^4}\left(\frac{\partial^4 r}{\partial \theta^4}+\frac{2}{\sin^2(\theta)}\frac{\partial^4 r}{\partial \theta^2 \partial \phi^2}+ \frac{1}{\sin^4(\theta)}\frac{\partial^4 r}{\partial \phi^4}\right)+F,
\end{eqnarray}
which again admits the solution $r(t)=R_0+Ft$. If we perform the linear stability analysis we arrive at the equation
\begin{eqnarray}
\nonumber
\frac{d\rho_{n,m}}{dt}=-\frac{K_0}{(Ft+R_0)^2}\left(n^2+ \frac{4}{3} m^2 \right)\rho_{n,m}- 2\frac{K_1}{(Ft+R_0)^3}\left(n^2+\frac{4}{3}m^2 \right) \rho_{n,m} \\
-\frac{K_2}{(Ft+R_0)^4}\left( n^4+\frac{8}{3}m^2n^2+\frac{8}{3}m^4 \right)\rho_{n,m},
\end{eqnarray}
that can be solved to yield
\begin{eqnarray}
\nonumber
\rho_{n,m}(t)=\exp \left(-\frac{K_2(8m^4+8m^2n^2+3n^4)}{9F}\left[\frac{1}{R_0^3}-\frac{1}{(R_0+Ft)^3}\right] \right. \\
-\frac{4m^2+3n^2}{3F} \left. \left[\frac{K_1}{R_0^2}-\frac{K_1}{(R_0+Ft)^2}+\frac{K_0}{R_0}-\frac{K_0}{R_0+Ft}\right]\right)\rho_{n,m}(0),
\label{sol3d}
\end{eqnarray}
and we see again that all the Fourier modes, except $m=n=0$, are stable for $t>0$. The case $m=n=0$ is marginal, but it is unimportant since it implies perturbations homogeneous in $\theta$ and $\phi$.

The effect of the $K_1$ term can be immediately understood by regarding Eqs.(\ref{sol2d}, \ref{sol3d}). It is a new mechanism for "dissipating" curvature. When a stochastic perturbation drives the solution away from the radially symmetric form, then this perturbation behaves as stated by this two equations. As can be clearly seen, the restoring of the symmetric form is faster in the $(2+1)-$dimensional case due to the presence of the $K_1$ term.

\section{Conclusions}

Motivated by the successful research on tumor growth by Br\'{u} {\it et al.}~\cite{bru1,bru2,bru3,bru4}, we have introduced theoretical models able to reproduce some of the features found in these experiments. We have also analyzed these models in order to better understand what is happening in the physical phenomenon. The models were derived as an expansion from a potential, which resulted a power series expansion in the mean curvature of the surface. The effect of the deterministic part of the equations is thus to reduce the mean curvature and its different powers, and this way to minimize the pressure in the tumor border and to favour tumor growth.

Since all the experiments were performed with two-dimensional tumors, it is important to develop theoretical models that represent what happens in these experiments, and to extend them to the three-dimensional case. This way we will get an insight of the evolution of the more realistic three-dimensional system, and we can derive conclusions that can be used as a guide for future experiments.

We have analyzed the effect of stochasticity in these equations, and we have shown that stochastic effects are much less relevant in the case of $(2+1)-$dimensional growth. Also, we have discussed the possible origin of the noise on the underlying disorder of the system, and how its intensity can be enhanced by increasing the number of inmune cells in
the tumor environment. This shows that the strategy of enhancing the inmune response in order to stop tumor growth should be less effective in the case of three-dimensional tumors. In addition to this, we have shown that a new term (the $K_1$ term) appeared with higher dimensionality. The origin of this term is very interesting, because it vanishes identically in one dimension for any geometry and in any dimension in the case of a planar geometry. Thus it appeared in the dynamics of the growing tumor as a combined effect of dimensionality and geometry. This term contributes to minimize the pressure and to favour tumor growth. Since it is present only in the three-dimensional case, it is another new mechanism that helps tumor propagation in this dimensionality, what lead us to conclude that it is much more difficult to stop a three-dimensional tumor than to stop its two-dimensional counterpart.

All along this work we have assumed that there are no overhangs in the interface in the radial direction, in such a way that a single valued solution of the corresponding stochastic growth equation makes sense for representing the interface evolution. There are, however, situations for which we cannot assume this. An interesting problem for future work is to derive a continuum model allowing overhangs and an arbitrary topology of the growing interface. Similar models were developed in a different context~\cite{keblinski1,keblinski2}, and allow a more detailed description of a complex growth phenomenon.

\section*{Acknowledgments}

This work has been partially supported by the Ministerio de Educaci\'{o}n y Ciencia (Spain) through Projects No. EX2005-0976 and FIS2005-01729.


\begin{thebibliography} {99}
\bibitem{polar} Our polar coordinates are, for $(1+1)-$dimensional geometry, $\vec{r}=(r(\theta)\cos(\theta),r(\theta)\sin(\theta))$, with $\theta \in [0,2\pi]$. For $(2+1)-$ dimensional geometry they read $\vec{r}=(r(\theta,\phi)\sin(\theta)\cos(\phi),r(\theta,\phi)\sin(\theta)\sin(\phi),r(\theta,\phi)\cos(\theta))$, with $\phi \in [0,2\pi]$ and $\theta \in [0,\pi]$.
\end{thebibliography}
\end{document}